\documentclass{aa}   
\usepackage[comma,authoryear]{natbib}
\usepackage{graphicx}
\usepackage[varg]{txfonts} 
\usepackage{longtable}

\newcommand{\eg}{{\it e.g.}}
\newcommand{\ie}{{\it i.e.}}
\newcommand{\ms}{m\,s$^{\rm -1}$}
\newcommand{\kms}{km\,s$^{\rm -1}$}
\newcommand{\msy}{m\,s$^{\rm -1}$\,y$^{\rm -1}$}
\newcommand{\msysq}{m\,s$^{\rm -1}$\,y$^{\rm -2}$}
\newcommand{\Mjup}{M$_{\rm Jup}$}

\newcommand{\ME}{M$_{\rm Earth}$}
\newcommand{\Msun}{M$_{\sun}$}
\newcommand{\Rsun}{R$_{\sun}$}
\newcommand{\vsini}{$v\sin{i}$}
\newcommand{\elodie}{E{\small LODIE}}
\newcommand{\sophie}{S{\small OPHIE}}
\newcommand{\sophiep}{S{\small OPHIE+}}
\newcommand{\harps}{H{\small ARPS}}

\begin{document}
   \title{Extrasolar planets and brown dwarfs around A--F type stars.
     \thanks{Based on observations made with the \sophie~spectrograph at the Observatoire de Haute-Provence (CNRS, France).}
     \thanks{Table 2 is only available in electronic form at the CDS via anonymous ftp to cdsarc.u-strasbg.fr (130.79.128.5) or via http://cdsweb.u-strabg.fr/cgi-bin/qcat?J/A+A/}}

   \subtitle{VIII. {\bf A giant planet orbiting the young star HD113337} }

   \author{
   S. Borgniet \inst{1} 
\and
   I. Boisse \inst{2}
\and
   A.-M. Lagrange \inst{1} 
\and 
   F. Bouchy \inst{3} 
\and
   L. Arnold\inst{4} 
\and
   R.F. D\'{\i}az\inst{5,6} 
\and
   F. Galland\inst{1}
\and 
   P. Delorme \inst{1}
\and
   G. H\'ebrard \inst{5,6}
\and
   A. Santerne\inst{2,3} 
\and
   D. Ehrenreich  \inst{7}
\and
   D. S\'egransan \inst{7} 
\and  
   X. Bonfils \inst{1} 
\and
   X. Delfosse \inst{1}  
\and
   N. C. Santos \inst{2,8} 
\and
   T. Forveille \inst{1} 
\and
   C. Moutou \inst{3} 
\and 
   S. Udry \inst{7} 
\and   
   A. Eggenberger \inst{7} 
\and 
   F. Pepe \inst{7}
\and
   N. Astudillo \inst{1}
\and
   G. Montagnier \inst{5,6}
   }

   \institute{
Institut de Plan\'etologie et d'Astrophysique de Grenoble, UMR5274 CNRS, Universit\'e Joseph Fourier, BP 53, 38041 Grenoble Cedex 9, France\\
     \email{Simon.Borgniet@obs.ujf-grenoble.fr}
\and
Centro de Astrofisica, Universidade do Porto, Rua das Estrelas, 4150-762 Porto, Portugal
\and
Aix Marseille Universit\'e, CNRS, LAM (Laboratoire d'Astrophysique de Marseille) UMR 7326, 13388, Marseille, France
\and
Aix Marseille Universit\'e, CNRS, OHP (Observatoire de Haute Provence), Institut Pyth\'eas, UMS 3470, 04870 Saint-Michel-l'Observatoire, France
\and
Institut d'Astrophysique de Paris, UMR7095 CNRS, Universit\'e Pierre \& Marie Curie, 98bis boulevard Arago, 75014 Paris, France
\and
Observatoire de Haute Provence, CNRS/OAMP, 04870 St Michel l' Observatoire, France 
\and
Observatoire de Gen\`eve, Universit\'e de Gen\`eve, 51 Chemin des Maillettes, 1290 Sauverny, Switzerland
\and
Departamento de Fisica e Astronomia, Faculdade de Ci\^encias,
Universidade do Porto, Rua do Campo Alegre, 4169-007 Porto, Portugal
}
   \date{Received ... ; accepted ...}

   % \abstract{}{}{}{}{}
   % 5 {} token are mandatory
   
   \abstract
   % context heading (optional)
   % {} leave it empty if necessary
   {}
   % aims heading (mandatory)
   {In the frame of the search for extrasolar planets and brown dwarfs around early-type main-sequence stars, we present the detection of a giant planet around  the young F-type star HD113337. We estimated the age of the system to be 150$_{-50}^{+100}$~Myr. Interestingly, an IR excess attributed to a cold debris disk was previously detected on this star.}
   % methods heading (mandatory)
   {The \sophie~ spectrograph on the 1.93m telescope at Observatoire de Haute-Provence was used to obtain  $\simeq$ 300 spectra over 6 years. We used our SAFIR tool, dedicated to the spectra analysis of A and F stars, to derive the radial velocity variations.}
   % results heading (mandatory)
   {The data reveal a $324.0^{+1.7}_{-3.3}$ days period that we attribute to a giant planet with a minimum mass of $2.83\pm 0.24$ \Mjup~in an eccentric orbit with $e=0.46\pm0.04$. A long-term quadratic drift, that we assign to be probably of stellar origin, is superimposed to the Keplerian solution.}
  % conclusions heading (optional), leave it empty if necessary {}
{}
   \keywords{techniques: radial velocities - stars: early-type - stars: planetary systems - stars: individual: HD113337}

   \maketitle
%
%________________________________________________________________

\section{Introduction}
Thanks to hundreds of planets discoveries \citep[http://exoplanet.eu/;][]{schneider11} mainly by radial velocity (RV) or transit surveys since \cite{mayor95}, our knowledge on exoplanets has dramatically improved. From these surveys, we now know that exoplanets are frequent around solar-type stars. More than 50$\%$ of these stars have planets with all kind of masses and, among them, about 14$\%$ have planets with masses larger than 50~\ME~\citep{mayor11}. An unexpected diversity of planet properties in separations, eccentricities and orbital motions (\eg~retrograde orbits) was revealed. These discoveries highlight the importance of dynamics, either through planet-planet or planet-disk interactions, in the building of the planet systems architectures. Meanwhile, many open questions remain regarding several aspects of planet formation, even in the case of giant planets. RV and transit explorations, after mainly characterizing Jupiter-like planets, are now detecting Neptune-like planets (10-40 \ME) and Super Earths (1.2-10 \ME). However, these techniques are still mostly limited to planets within a separation of typically 5 AU from their parent stars. Moreover, they target generally solar-type main sequence stars or evolved stars. Imaging techniques are sensitive only to giant planets orbiting at large separations, from 8 to 1000AU, around young stars, typically $\simeq$1-100 Myr. These limitations have several consequences. In particular, it is difficult to test the impact of the host star mass on the planets properties, while formation models predict different behaviors according to the stellar mass \citep[roughly, for the core-accretion scenario, higher mass planets are expected around more massive stars,][]{kennedy08,mordasini09}. RV and microlensing surveys revealed so far very few massive planets around M-stars \citep[see \eg,][]{batista11}, but these low-mass stars show an abundance of light planets at short distance \citep{bonfils12}. At the other end of the stellar mass spectrum, several giant planets have been found around massive, evolved stars at orbital distances typically greater than 0.7\,AU \citep[see \eg,][and references there-in]{johnson11}. It is not yet clear how the stellar evolution impacts the observed planetary distributions and properties. The fact that transit surveys detect planets at very short periods seems to confirm an evolutionary effect. Close/intermediate separations have then to be investigated by observing massive main-sequence stars. Massive, early-type main-sequence stars show few spectral lines that are in addition generally broadened by a high stellar rotation rate \citep[][hereafter Paper\,I]{galland05a}. Classical techniques used for solar-type stars, such as masking techniques, are therefore not adapted to these stars. In this frame, we developed the SAFIR tool, dedicated to the measurement of the RV in spectra of A--F type stars. SAFIR is described in Paper I and is based on the Fourier interspectrum method \citep{chelli00}. We initiated a survey dedicated to the search for extrasolar planets and brown dwarfs around a volume-limited sample of A--F main-sequence stars {\it i)} with the \elodie~fiber-fed echelle spectrograph \citep{baranne96} mounted on the 1.93-m telescope at the Observatoire de Haute-Provence (OHP, France) in the northern hemisphere, and then with its successor \sophie~\citep{bouchy06}, and {\it ii)} with the \harps~spectrograph \citep{pepe02} mounted on the 3.6-m ESO telescope at La Silla Observatory (Chile) in the southern hemisphere. A few giant planets or planet candidates were reported around these targets \citep{galland05a,galland06,desort08,desort09,lagrange12a}.

 We present here the detection of a giant planet around HD113337 observed in the course of our SP4 program with the \sophie~Consortium \citep{bouchy09}. We present the stellar properties of HD113337 in Sect. 2, and the \sophie~data in Sect. 3. In Sect. 4, we discuss the origin of the observed RV variations before concluding in Sect. 5.

%__________________________________________________________________

\section{Stellar characteristics} \label{stellar_param}

\subsection{General properties}

 HD113337 (HIP63584, HR4934) is a bright F6V star \citep{hoffleit91}, located at $36.9\pm0.4$\,pc from the Sun \citep{vanleeuwen07}. The main stellar parameters are reported in Table~\ref{param_s}. \cite{rhee07} estimated a radius of $1.5$~\Rsun~from the stellar spectral energy distribution and the parallax. \cite{reid07} identified an M4 star companion (2MASS J13013268+6337496) at 120~arcsecs ($\geq$ 4000AU). This companion is associated to an X-ray emission detected with ROSAT \citep{haakonsen09}. An IR excess  was detected by IRAS that \cite{rhee07} attributed to a cold dust of $\simeq100$K in a ring at $\simeq18$~AU from the star. On the other hand, using Spitzer data, \cite{moor11} estimated that the dust is located at $55\pm 3$~AU and has a temperature of $53\pm1$~K. 

\subsection{Age of the star}

Several approaches to estimate the age of the star have led to different and initially incompatible results. The Geneva catalog assigns an age of 1.5\,Gyr to HD113337 \citep{holmberg09}, based on the Padova stellar evolution model \citep{holmberg07}. However, based on this model, most of the members of young associations (\eg~the Beta Pic moving group) are also assigned an age of 1-2\,Gyr, much older than their actual age ($<100$ Myr). The same discrepancy being possible for HD113337, we therefore do not rely on the age estimation based on this model.
 F-type stars that have an effective temperature around 6600\,K are known to present a lithium (Li) gap that depends on the age \citep[see \eg,][]{boesgaard87}. Our star's effective temperature is estimated to be 6545\,K \citep{boesgaard86}. We therefore calculated its Li abundance and used the age-dependent relation derived by \cite{boesgaard87} to estimate its age. A look at \sophie~high signal-to-noise ratio (SNR) spectrum of the primary star shows no sign of the signature of the lithium line at 6707.8\AA. The analysis suggests an upper value for the equivalent width of this line to be 1m\AA, that corresponds to an upper limit for the Li abundance in HD113337 of A(Li)=1.5\,dex. This value was derived using the radiative transfer code MOOG \citep{sneden73}, and a grid of Kurucz ATLAS\,9 model atmospheres \citep{kurucz93}. The input effective temperature and metallicity are the same as presented in Table\,1.
 We compared the derived Li abundance with those of other Li-gap stars in the open cluster M35 \citep{steinhauer04}. HD113337 abundance value is lower than any of the Li abundances observed in this 160$\pm$20\,Myr cluster. Although some upper limit Li abundances are observed by \cite{steinhauer04}, this result suggests that HD113337 is likely older than $\simeq$150\,Myr. Its Li abundance is however compatible with those observed in the older 700\,Myr old Hyades cluster \citep{steinhauer04}. We note that Li abundances are most sensitive to the effective temperature \citep[see \eg,][]{israelian04}. However, adopting a slighly different value, for example by 100K, will only change the derived Li abundance by 0.08 dex.

\renewcommand{\arraystretch}{1.25}
\begin{table}[t!]
\caption{HD113337 stellar properties.}  
\label{param_s}
\begin{center}
\begin{tabular}{l l c}\\
        \hline
      	\hline
        Parameter        &                      & HD113337  \\
	\hline
        Spectral Type    &                      & F6V\tablefootmark{a} \\
        $B-V$            &                      & 0.43\tablefootmark{b}\\
        $V$              &                      & 6.0\tablefootmark{b}\\
        \vsini           & [\kms]               & 6.1\tablefootmark{c}\\ 
         \vsini        &  [\kms]          & $6.3 \pm 1$\tablefootmark{d} \\
        $\pi$            & [mas]                &  $27.11 \pm 0.29$ \tablefootmark{e}\\
	$[$Fe/H$]$       &                      & $0.07$ \tablefootmark{a} \\
        log($T_{\rm eff}$) & [K]                  & $3.818 \pm 0.01$\tablefootmark{a} \\
        $\log{g}$        & [dex]                & $4.21 \pm 0.08$\tablefootmark{f}  \\
        Mass             & [\Msun]              & $1.40 \pm 0.14$\tablefootmark{f} \\
        Radius           & [\Rsun]              & $1.50 \pm 0.15$\tablefootmark{g}\\
	Age              & [Myr]                & 150$_{-50}^{+100}$\\
        \hline
\end{tabular}
\tablefoot{
\tablefoottext{a}{\cite{hoffleit91}}
\tablefoottext{b}{\cite{perryman97}}
\tablefoottext{c}{estimation from the SAFIR software}
\tablefoottext{d}{ estimation from the SOPHIE data reduction system \citep{boisse10b}}
\tablefoottext{e}{\cite{vanleeuwen07}} 
\tablefoottext{f}{\cite{allende99}}
\tablefoottext{g}{\cite{rhee07}}
}
\end{center}
\end{table}
\renewcommand{\arraystretch}{1}

 On the other hand, we can determine the age from an analysis of the M dwarf companion detected by \cite{reid07}. \cite{rhee07} estimated the age of the companion to be about 50~Myr based on its ($K$,$V-K_S$) properties and showed that the galactic UVW motion of the system is typical of a young population. We did a more robust isochronal analysis, using absolute $K_S$-band magnitude and a spectroscopic determination of the spectral type by \cite{moor11}.  We converged to a similar age of 40$\pm$20~Myr, also incompatible with the Li age determination of the primary. This discrepancy could put into question that the M star is actually a companion to \object{HD\,113337}. But, a recent lucky-imaging study by \cite{janson12} offered a way to reconcile the age estimations. These authors observed that the companion is a close, moderate brightness contrast M-type binary. Binarity affects the isochronal age estimation in two different ways that both work towards an older age. First the actual absolute magnitude of the main component is fainter by ~0.4 magnitude in $K_S$-band when accounting for the flux coming from the secondary component. Second, the presence of a cooler companion biases the spectral type determination using unresolved spectra towards cooler temperature estimates. When accounting for both effects, we find that the primary component of the M dwarf companion has a temperature of 3350$\pm$100~K and an absolute $K_S$ magnitude of 7.0. Using the BT-Settl isochrones of \cite{allard12} derived from the stellar evolution models of \cite{chabrier00}, we find an age of 100$_{-50}^{+100}$~Myr.
We note that our temperature estimate includes larger error bars than the estimation of \cite{moor11} which had an uncertainty of 70~K. The reason is that we corrected the systematic error in effective temperature arising from binarity. This correction is quite crude because we can only use one single low SNR point of resolved photometry \citep[$\Delta~z$'=0.9$\pm$0.27,][]{janson12}. A solution to increase our accuracy  would be to obtain resolved spectroscopy. Moreover, it would permit to establish whether the M-dwarf system is only a binary, or contains more components.

Finally, using the constraints from both analysis, we adopt the most probable age of 150$_{-50}^{+100}$~Myr for the HD113337AB system.

  %---------------------------------
  \section{Spectroscopic data}
  %---------------------------------
  \subsection{Description of the observations }

 We obtained 312 high signal-to-noise ratio (SNR) spectra with \sophie, in the 3872--6943\,\AA$ $ range, in high-resolution mode ($R\approx75\,000$ at 550 nm). The time span of the data is 2193 days, between Jan. 2007 and Jan. 2013, but all the data except three were taken after Feb. 2008. The exposure times, typically between 180 and 500s, were adapted to obtain an average SNR of $155$. The exposures were performed with simultaneous-thorium spectra to follow and correct for the drift of the instrument due to local temperature and pressure variations.

  The RVs are computed using SAFIR. We selected spectra resting on two criterions:
\begin{itemize}
\item first, the SNR has to be greater than $80$;
\item second, the atmospheric absorption has to be lower than 3. For each spectrum, the absorption corresponds to the deviation between the stellar apparent magnitude in the V band and an empirical magnitude derived from the exposure time and the SNR.
\end{itemize}
 We thus ended up with 266 RV values. As HD113337 is an F6 star, its spectra contain sufficient lines for the \sophie~automatic data reduction software (DRS) to derive RVs \citep{bouchy09}. We verified that DRS RV values are consistent with the SAFIR ones.  
Before June 2011 an instrumental effect due to the insufficient scrambling of one multimode fiber that led to non-uniform illumination of the entrance of the spectrograph \citep{boisse10a} was observed in the data (later referred as \sophie~data). This effect was significantly decreased thanks to a fiber-link modification, which includes a piece of octagonal-section fiber \citep[see][later referred as \sophiep~data]{perruchot11,bouchy13}. We adapted to SAFIR the method of \cite{diaz12} to correct the \sophie~data.
  The RVs quadratic mean is 57.1\,\ms, divided in 57.8\,\ms~and 43.8\,\ms~for \sophie~and \sophiep~values, respectively. Accounting  for 3.2\,\ms~of photon noise and 5 \,\ms~of assumed instrumental stability, the RV uncertainty is 8.2\,\ms~on average for \sophiep~values. The error bars on the \sophie~RVs also take into account the correction applied, and their average uncertainty is 12.4\,\ms. The RV measurements of HD113337 are listed in Table~\ref{rv} and are available at the Strasbourg astronomical Data Center (CDS). It contains in its Cols. 1-4, the time of the observation (barycentric Julian date), the RV, its error, and a flag to distinguish \sophie~and \sophiep~values, respectively.

\onllongtab{2}{
\begin{longtable}{lllc}
\caption{HD113337 radial velocities.}
\label{rv}\\
\hline
\hline 
Date of     &   RV    & RV & Period Flag\\
observation &         & uncertainty & (0: SOPHIE) \\
JD-2454000 &  \kms &  \kms  &  (1: SOPHIE+) \\
\hline 
126.71449	&	-0.2596167	&	0.0153933	&	0	\\
127.67123	&	-0.2225158	&	0.0105579	&	0	\\
127.68597	&	-0.2298054	&	0.0099728	&	0	\\
524.63478	&	-0.0389689	&	0.0121061	&	0	\\
536.64281	&	-0.0055641	&	0.0151627	&	0	\\
543.56786	&	-0.0256958	&	0.0122984	&	0	\\
543.57458	&	-0.0180662	&	0.0124291	&	0	\\
549.54104	&	-0.016673	&	0.0121501	&	0	\\
549.54906	&	-0.0274646	&	0.0122232	&	0	\\
584.53667	&	-0.0061845	&	0.0130287	&	0	\\
593.47478	&	-0.0158597	&	0.0147139	&	0	\\
593.47865	&	-0.0068883	&	0.0116872	&	0	\\
594.44039	&	-0.0665764	&	0.0119836	&	0	\\
594.44348	&	-0.0636492	&	0.0119661	&	0	\\
668.34569	&	-0.0224724	&	0.0119584	&	0	\\
668.34882	&	-0.0336648	&	0.012027	&	0	\\
822.67439	&	-0.0582648	&	0.0125285	&	0	\\
822.68001	&	-0.0455546	&	0.0130484	&	0	\\
852.63289	&	0.0217796	&	0.0144921	&	0	\\
852.6368	&	0.0366221	&	0.0145534	&	0	\\
852.70044	&	0.0338246	&	0.0147837	&	0	\\
852.70435	&	0.0230074	&	0.0147639	&	0	\\
853.69802	&	-0.0153359	&	0.014974	&	0	\\
853.70093	&	-0.0092295	&	0.0151613	&	0	\\
872.54037	&	0.0310181	&	0.0154554	&	0	\\
872.54765	&	0.0270524	&	0.0135919	&	0	\\
879.61365	&	0.0607119	&	0.013814	&	0	\\
879.61669	&	0.054122	&	0.0129526	&	0	\\
881.65862	&	0.0350246	&	0.011487	&	0	\\
881.66935	&	0.034134	&	0.013855	&	0	\\
882.63619	&	0.0510694	&	0.0113607	&	0	\\
882.64255	&	0.0630943	&	0.012005	&	0	\\
883.68179	&	0.0426047	&	0.0117156	&	0	\\
883.68984	&	0.0375659	&	0.0113528	&	0	\\
884.56559	&	0.063648	&	0.0122771	&	0	\\
884.56999	&	0.0598469	&	0.012077	&	0	\\
885.55622	&	0.0719913	&	0.0128611	&	0	\\
885.56428	&	0.0597266	&	0.0111951	&	0	\\
890.62201	&	0.0091506	&	0.0122064	&	0	\\
890.62896	&	0.019155	&	0.0116503	&	0	\\
902.51826	&	0.0450805	&	0.0161627	&	0	\\
902.52116	&	0.046029	&	0.015023	&	0	\\
904.48591	&	0.0366957	&	0.0138832	&	0	\\
904.4903	&	0.037686	&	0.0143765	&	0	\\
906.53037	&	0.0668746	&	0.011008	&	0	\\
906.53841	&	0.0517159	&	0.0115312	&	0	\\
911.56387	&	0.0096629	&	0.0122362	&	0	\\
911.57465	&	0.0164869	&	0.011963	&	0	\\
913.48494	&	0.0418106	&	0.0121906	&	0	\\
913.4905	&	0.059651	&	0.0116508	&	0	\\
915.6173	&	0.0353425	&	0.0164099	&	0	\\
915.62175	&	0.0227972	&	0.0171932	&	0	\\
925.55332	&	0.0249043	&	0.0116172	&	0	\\
925.55676	&	0.0174431	&	0.011656	&	0	\\
926.56987	&	0.0161248	&	0.0145417	&	0	\\
926.57277	&	0.0305184	&	0.0146176	&	0	\\
934.4908	&	0.03333	        &	0.0122084	&	0	\\
934.49372	&	0.0379068	&	0.0122182	&	0	\\
936.4311	&	0.0364464	&	0.0126999	&	0	\\
941.5236	&	0.0113076	&	0.0121404	&	0	\\
941.5309	&	0.015826	&	0.0119785	&	0	\\
951.43237	&	0.0128217	&	0.0107996	&	0	\\
951.43746	&	0.014196	&	0.0110105	&	0	\\
952.43224	&	0.0425062	&	0.0129868	&	0	\\
952.43515	&	0.0572271	&	0.0132158	&	0	\\
953.38015	&	0.0277884	&	0.0125816	&	0	\\
953.38304	&	0.0238852	&	0.0126013	&	0	\\
958.39955	&	0.0707394	&	0.0125476	&	0	\\
958.40243	&	0.0770117	&	0.0128296	&	0	\\
971.36169	&	0.0082993	&	0.0121765	&	0	\\
971.36459	&	0.0217457	&	0.0121721	&	0	\\
1006.3823	&	0.0409544	&	0.0128717	&	0	\\
1006.3852	&	0.0579549	&	0.0130578	&	0	\\
1012.3658	&	-0.057519	&	0.0113091	&	0	\\
1029.3632	&	-0.0510263	&	0.0125147	&	0	\\
1029.3662	&	-0.051158	&	0.0126329	&	0	\\
1031.3606	&	-0.0385713	&	0.0164968	&	0	\\
1031.3648	&	-0.0388165	&	0.0167477	&	0	\\
1053.3419	&	-0.094407	&	0.0133913	&	0	\\
1053.3527	&	-0.0710718	&	0.0115508	&	0	\\
1058.3387	&	-0.0679094	&	0.0117131	&	0	\\
1058.346	&	-0.0656624	&	0.0118423	&	0	\\
1238.7008	&	0.0989002	&	0.0134204	&	0	\\
1238.7081	&	0.0759548	&	0.0140823	&	0	\\
1267.5764	&	0.0300148	&	0.0124577	&	0	\\
1267.5819	&	0.0353871	&	0.0118056	&	0	\\
1269.5594	&	0.0462886	&	0.0106449	&	0	\\
1269.5623	&	0.0432161	&	0.0106735	&	0	\\
1270.52	        &	0.0567584	&	0.0153135	&	0	\\
1270.5229	&	0.0539644	&	0.0150571	&	0	\\
1282.5496	&	0.0473729	&	0.0126118	&	0	\\
1282.5525	&	0.0523341	&	0.0129308	&	0	\\
1284.5731	&	0.0273371	&	0.0108279	&	0	\\
1284.576	&	0.0081573	&	0.0108278	&	0	\\
1286.4776	&	0.0510374	&	0.0104482	&	0	\\
1286.4843	&	0.0716583	&	0.0104424	&	0	\\
1288.4863	&	0.0493318	&	0.0115476	&	0	\\
1288.4896	&	0.0448531	&	0.0111833	&	0	\\
1292.4539	&	-0.0019308	&	0.0114981	&	0	\\
1292.4565	&	0.0021357	&	0.0114416	&	0	\\
1295.4996	&	0.0671524	&	0.0102579	&	0	\\
1295.5034	&	0.0735785	&	0.0102547	&	0	\\
1316.462	&	0.0169111	&	0.0110343	&	0	\\
1316.4663	&	0.0227399	&	0.0110413	&	0	\\
1324.5011	&	0.0013523	&	0.0122251	&	0	\\
1324.5056	&	0.0164634	&	0.0148875	&	0	\\
1330.3752	&	0.0483305	&	0.0128452	&	0	\\
1330.3781	&	0.0387688	&	0.0139012	&	0	\\
1344.3962	&	0.0128634	&	0.011529	&	0	\\
1344.3991	&	0.0111788	&	0.0117061	&	0	\\
1346.3494	&	-0.0393047	&	0.0141665	&	0	\\
1346.3523	&	-0.007524	&	0.0134268	&	0	\\
1361.3642	&	-0.013428	&	0.0107473	&	0	\\
1367.3946	&	-0.0383995	&	0.0132958	&	0	\\
1382.3636	&	-0.0445566	&	0.0119291	&	0	\\
1382.3665	&	-0.0357518	&	0.0120646	&	0	\\
1383.3715	&	-0.048996	&	0.013358	&	0	\\
1383.3745	&	-0.0441935	&	0.0135168	&	0	\\
1392.3596	&	-0.0749329	&	0.0121567	&	0	\\
1395.3515	&	-0.0830119	&	0.0136584	&	0	\\
1395.3544	&	-0.0777054	&	0.0128558	&	0	\\
1397.3438	&	-0.0404038	&	0.0127475	&	0	\\
1397.3468	&	-0.0212646	&	0.0128647	&	0	\\
1399.3546	&	-0.0495331	&	0.0105173	&	0	\\
1399.359	&	-0.0530482	&	0.0104723	&	0	\\
1401.3688	&	-0.0374041	&	0.013479	&	0	\\
1401.3761	&	-0.0653215	&	0.0144244	&	0	\\
1409.348	&	-0.0660864	&	0.0121639	&	0	\\
1409.3509	&	-0.0554636	&	0.0120011	&	0	\\
1498.2452	&	0.1039677	&	0.0118739	&	0	\\
1503.229	&	0.0670044	&	0.0105982	&	0	\\
1503.2351	&	0.0649353	&	0.0111255	&	0	\\
1557.7089	&	0.0672138	&	0.0179282	&	0	\\
1563.728	&	0.0826305	&	0.0126138	&	0	\\
1579.6829	&	0.0478971	&	0.0111125	&	0	\\
1579.6859	&	0.0570132	&	0.0109404	&	0	\\
1584.6415	&	0.0942919	&	0.0122702	&	0	\\
1584.6457	&	0.0977787	&	0.0116229	&	0	\\
1585.6586	&	0.0597548	&	0.0111127	&	0	\\
1585.6808	&	0.0419338	&	0.0111421	&	0	\\
1587.6931	&	0.0443352	&	0.0115056	&	0	\\
1587.6986	&	0.0499658	&	0.0111196	&	0	\\
1619.5325	&	0.0179725	&	0.0109274	&	0	\\
1619.5381	&	0.0283572	&	0.0109833	&	0	\\
1627.6883	&	0.023539	&	0.011277	&	0	\\
1627.6914	&	0.0275496	&	0.0112617	&	0	\\
1629.5694	&	0.0197756	&	0.0100076	&	0	\\
1629.5768	&	0.0240008	&	0.0106935	&	0	\\
1631.4752	&	0.0015472	&	0.0123249	&	0	\\
1631.4802	&	0.0184798	&	0.0126132	&	0	\\
1638.5201	&	0.0179916	&	0.0120943	&	0	\\
1638.5309	&	0.0168867	&	0.0132883	&	0	\\
1640.5795	&	0.0453646	&	0.0109411	&	0	\\
1640.5839	&	0.0374772	&	0.0110087	&	0	\\
1641.5369	&	-0.0084004	&	0.0104148	&	0	\\
1641.5425	&	-0.0060748	&	0.0103176	&	0	\\
1659.4823	&	-0.0350745	&	0.011631	&	0	\\
1659.4853	&	-0.0362565	&	0.0125192	&	0	\\
1660.3919	&	0.0059588	&	0.0120043	&	0	\\
1660.3948	&	0.0027584	&	0.0119655	&	0	\\
1662.529	&	-0.0127341	&	0.0118991	&	0	\\
1662.5319	&	-0.0056402	&	0.0118764	&	0	\\
1669.4128	&	-0.0270591	&	0.0131353	&	0	\\
1669.4158	&	-0.0230812	&	0.0129508	&	0	\\
1671.4091	&	-0.0177911	&	0.012751	&	0	\\
1671.412	&	-0.0264192	&	0.0129996	&	0	\\
1678.3637	&	-0.0098259	&	0.0137968	&	0	\\
1681.362	&	-0.0153006	&	0.0117559	&	0	\\
1681.3653	&	-0.0260836	&	0.0113727	&	0	\\
1682.4537	&	-0.0429714	&	0.0117408	&	0	\\
1682.4593	&	-0.0611066	&	0.0127487	&	0	\\
1685.4624	&	-0.02706	&	0.0113371	&	0	\\
1685.4692	&	-0.0227569	&	0.0109892	&	0	\\
1686.4591	&	-0.075625	&	0.0112994	&	0	\\
1686.4633	&	-0.07707	&	0.0117506	&	0	\\
1694.4875	&	-0.0868201	&	0.0130581	&	0	\\
1694.4908	&	-0.0805795	&	0.0133028	&	0	\\
1698.4203	&	-0.0882925	&	0.0120913	&	0	\\
1698.4233	&	-0.1117705	&	0.0119962	&	0	\\
1699.3612	&	-0.0713517	&	0.0126621	&	0	\\
1699.3641	&	-0.0899508	&	0.0120793	&	0	\\
1702.3531	&	-0.110292	&	0.0100354	&	0	\\
1702.3573	&	-0.0937016	&	0.0100651	&	0	\\
1704.4133	&	-0.0834264	&	0.0113492	&	0	\\
1704.4171	&	-0.0880817	&	0.0113746	&	0	\\
1706.3762	&	-0.0787974	&	0.0115273	&	0	\\
1706.3797	&	-0.0924939	&	0.0116714	&	0	\\
1756.3565	&	-0.051511	&	0.00985	        &	1	\\
1756.3594	&	-0.045695	&	0.009682	&	1	\\
1758.3457	&	-0.0466	        &	0.007994	&	1	\\
1758.349	&	-0.051098	&	0.008035	&	1	\\
1762.4088	&	-0.030175	&	0.009685	&	1	\\
1762.4121	&	-0.038849	&	0.009402	&	1	\\
1785.3224	&	0.012936	&	0.008697	&	1	\\
1785.3253	&	0.012231	&	0.008725	&	1	\\
1796.3177	&	0.04169	        &	0.008272	&	1	\\
1901.7161	&	0.051405	&	0.007796	&	1	\\
1901.7212	&	0.048541	&	0.007807	&	1	\\
1906.6979	&	0.020837	&	0.010047	&	1	\\
1906.7004	&	0.020973	&	0.00887	        &	1	\\
1906.7039	&	0.01496	        &	0.008802	&	1	\\
1906.7064	&	0.011338	&	0.008841	&	1	\\
1930.6899	&	0.02133	        &	0.010006	&	1	\\
1934.7266	&	-0.001449	&	0.009231	&	1	\\
1934.7296	&	-0.004032	&	0.009507	&	1	\\
1963.6566	&	-0.025239	&	0.009322	&	1	\\
1963.6608	&	-0.027487	&	0.009709	&	1	\\
1981.5345	&	-0.068348	&	0.007559	&	1	\\
1981.5375	&	-0.066563	&	0.007549	&	1	\\
1982.5529	&	-0.043742	&	0.007532	&	1	\\
1982.5558	&	-0.039288	&	0.007536	&	1	\\
1984.4905	&	-0.020948	&	0.00844 	&	1	\\
1984.4986	&	-0.025738	&	0.007749	&	1	\\
1995.6008	&	-0.03791	&	0.007619	&	1	\\
1995.6057	&	-0.037477	&	0.007699	&	1	\\
1996.562	&	-0.054647	&	0.008197	&	1	\\
1996.5649	&	-0.053332	&	0.008107	&	1	\\
1998.6163	&	-0.029189	&	0.007863	&	1	\\
1998.6196	&	-0.026745	&	0.007897	&	1	\\
2018.4838	&	-0.05546	&	0.00731 	&	1	\\
2018.4922	&	-0.046048	&	0.00752 	&	1	\\
2019.4596	&	-0.050644	&	0.008063	&	1	\\
2019.4634	&	-0.058727	&	0.008022	&	1	\\
2026.4421	&	-0.054393	&	0.007671	&	1	\\
2026.4495	&	-0.056232	&	0.007906	&	1	\\
2027.5577	&	-0.043676	&	0.008045	&	1	\\
2027.5606	&	-0.043018	&	0.008092	&	1	\\
2029.5431	&	-0.053642	&	0.007951	&	1	\\
2029.5469	&	-0.058752	&	0.007932	&	1	\\
2043.4663	&	-0.109213	&	0.008561	&	1	\\
2043.4712	&	-0.097366	&	0.008237	&	1	\\
2044.5994	&	-0.081487	&	0.009118	&	1	\\
2050.4761	&	-0.136944	&	0.007081	&	1	\\
2050.4846	&	-0.146212	&	0.007678	&	1	\\
2050.4884	&	-0.142721	&	0.007704	&	1	\\
2079.4877	&	-0.095869	&	0.008347	&	1	\\
2079.4926	&	-0.0971 	&	0.008644	&	1	\\
2106.3847	&	-0.038415	&	0.007722	&	1	\\
2106.3879	&	-0.03651	&	0.007783	&	1	\\
2109.4098	&	0.010932	&	0.007777	&	1	\\
2109.4149	&	0.013416	&	0.007887	&	1	\\
2117.3609	&	0.006159	&	0.008002	&	1	\\
2117.3638	&	0.014303	&	0.00807	&	1	\\
2118.3685	&	-0.017431	&	0.007982	&	1	\\
2118.3714	&	-0.020039	&	0.008045	&	1	\\
2120.3517	&	0.00943	        &	0.008179	&	1	\\
2120.3546	&	0.01021	        &	0.008153	&	1	\\
2137.3487	&	-0.02078	&	0.007059	&	1	\\
2139.3716	&	-0.013609	&	0.007915	&	1	\\
2140.3568	&	-0.012979	&	0.007513	&	1	\\
2230.2358	&	-0.022836	&	0.008152	&	1	\\
2230.2387	&	-0.021532	&	0.008337	&	1	\\
2288.683	&	-0.065531	&	0.007763	&	1	\\
2288.6892	&	-0.065992	&	0.008496	&	1	\\
2289.7346	&	-0.048047	&	0.009239	&	1	\\
2289.738	&	-0.045157	&	0.009032	&	1	\\
2291.7296	&	-0.07685	&	0.007755	&	1	\\
2291.7335	&	-0.075707	&	0.007637	&	1	\\
2296.6789	&	-0.052402	&	0.007641	&	1	\\
2296.6821	&	-0.057838	&	0.007919	&	1	\\
2314.6812	&	-0.091407	&	0.007693	&	1	\\
2314.6863	&	-0.091224	&	0.008491	&	1	\\
2316.696	&	-0.119523	&	0.007581	&	1	\\
2316.6999	&	-0.117391	&	0.007651	&	1	\\
2319.6936	&	-0.101297	&	0.008156	&	1	\\
2319.6985	&	-0.102347	&	0.008519	&	1	\\
\hline
\end{longtable}
}

\subsection{Radial velocity variations }     

 \begin{figure}[t!] 
 \centering
 \includegraphics[width=0.7\hsize]{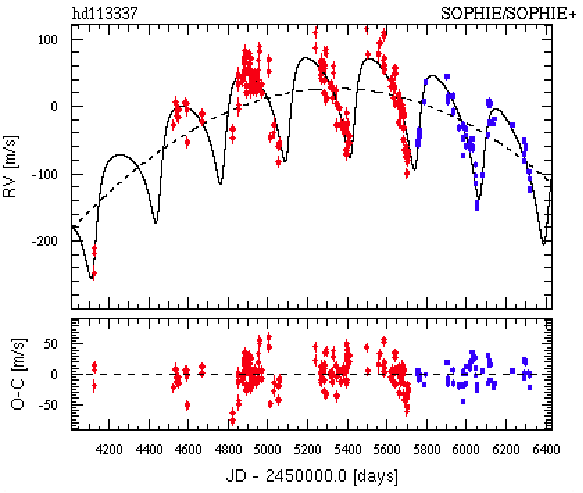}
 \caption{\textit{Top:} Keplerian fit of HD113337 RV variations with a planet plus a quadratic law (\sophie\,and \sophiep\,values are plotted in red and blue, respectively). \textit{Bottom:} Residuals of the fit as a function of time.}
 \label{fit_2CC_all}
 \end{figure}

\begin{figure}[t!]
  \centering
  \includegraphics[width=0.9\hsize]{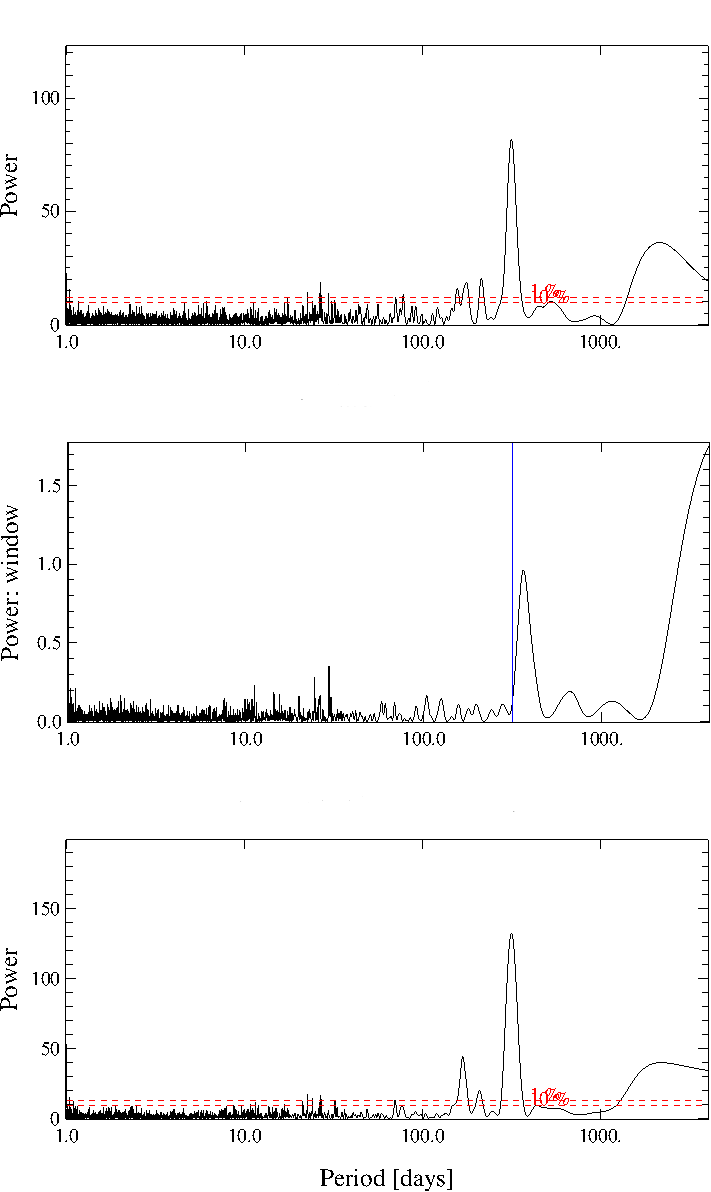}
  \caption{\textit{Top:} Periodogram of HD133337 RVs. The false alarm probabilities at 1 and 10\% are drawn in red dashed lines. \textit{Middle:} Periodogram of the temporal sampling of the data. The 316-days period is indicated by a blue line. Note the difference in the $y$-scale. \textit{Bottom:} Periodogram of simulated RVs of a star hosting a planet on a circular orbit with a 316 days period. The temporal sampling is the same than for HD113337 data.}
  \label{rv_perio}
  \end{figure}

The RV data are displayed in Fig.\,\ref{fit_2CC_all}. They show a periodic signal with a peak-to-peak amplitude of about 300\,\ms~over the whole period of time considered, and of about 200\,\ms~ if we consider only the data recorded in 2008 and later. Such amplitudes are much larger than the uncertainties ($\simeq$10\,\ms~in average, see above). The Lomb-Scargle periodogram of the RV variations is given in Fig.\,\ref{rv_perio}. It is calculated by SAFIR through an adapted version of the CLEAN algorithm \citep{roberts87} and gives the normalized power spectrum of the RV data (\ie, the square modulus of the RV data Fourier Transform) versus the period range. The significance of the periodogram peaks is tested thanks to the False Alarm Probabilities (FAP). The FAP are estimated using a bootstrapping approach, where the RV data are randomly shuffled and the corresponding periodograms are calculated. The highest peak is at a period of $\simeq$ 316 days (power = 82), which we will attribute to a planet (see below). A peak can also be found at about 2146 days (power = 36), and some other peaks with a FAP smaller than 1\%, at 214 days (power = 20), 176 days (power = 18) and 157 days (power = 16), the latter being an alias of the planet period. Other smaller peaks between 10 and 80 days can be attributed to the temporal sampling. For comparison, the periodogram of the window is also given, as well as an example of a periodogram that would be induced by a planet on a circular orbit with a period of 316 days. As expected, apart from the peak at the planet period, several other peaks, due to the temporal sampling are also observed at periods similar to the observed ones.

\subsection{Line profile variations}
The SOPHIE automatic pipeline cross-correlates the spectra with a G2-type mask. This is relevant as this spectral type is near that of HD113337. Each resulting cross-correlation function (CCF) is fitted by a Gaussian. Two parameters of the CCF, the bisector velocity span (BIS) and the full width half maximum (FWHM), allow the monitoring of line profile deformations, that could induce RV variations not related to a Doppler shift due to orbital motion. 

\begin{figure}[t!]
\centering
\includegraphics[width=0.8\hsize]{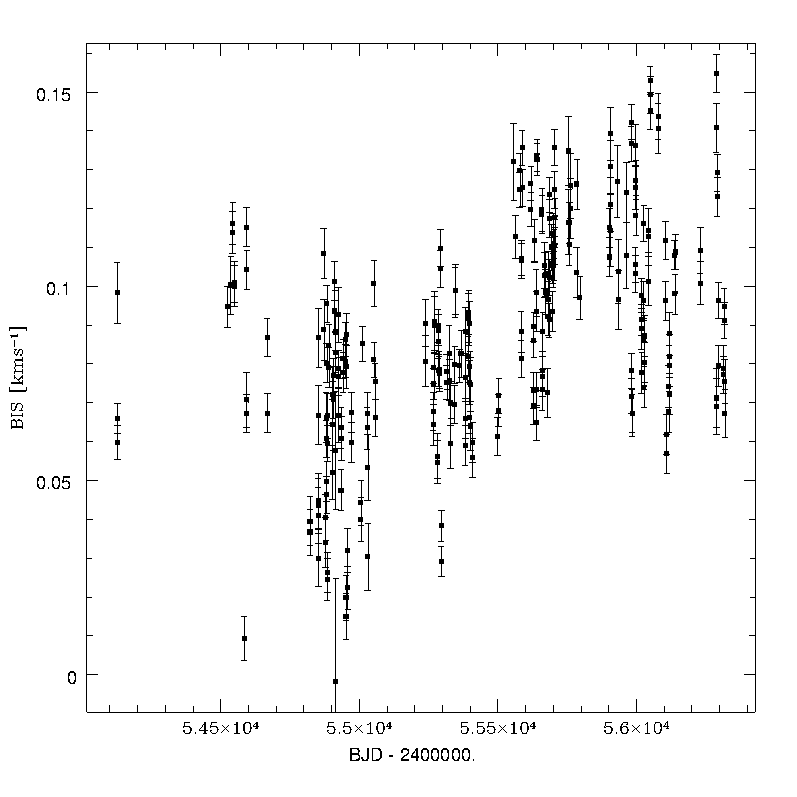}
\includegraphics[width=0.8\hsize]{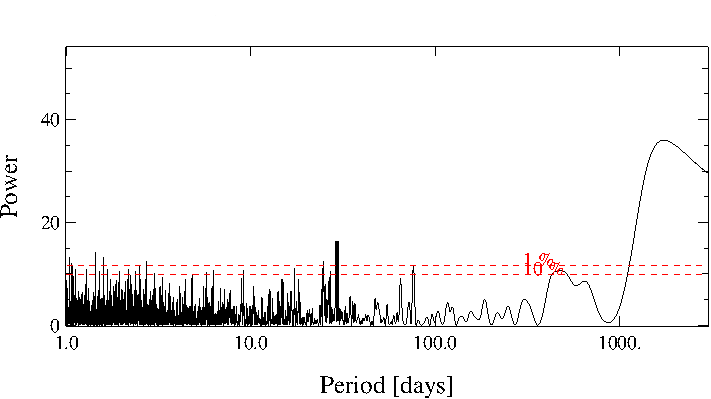}
\caption{\textit{Top:} BIS versus time.  \textit{Bottom:}  Periodogram of the BIS variations.}
\label{bis_span}
 \end{figure}

 We show in Figure~\ref{bis_span} the BIS as a function of time and its Lomb-Scargle periodogram, together with values of FAP. No BIS temporal variations are seen with a period of $\simeq$300 days, and the periodogram does not show any peak at 300-400 days. But, a long-term low amplitude BIS variation could be seen, highlighted by the highest peak in the periodogram at a period greater than 1000~days. No significant correlation is seen between RVs and BISs (\ie\, Pearson's correlation coefficient $<\,0.4$).

\begin{figure}[t!]
    \centering
     \includegraphics[width=0.8\hsize]{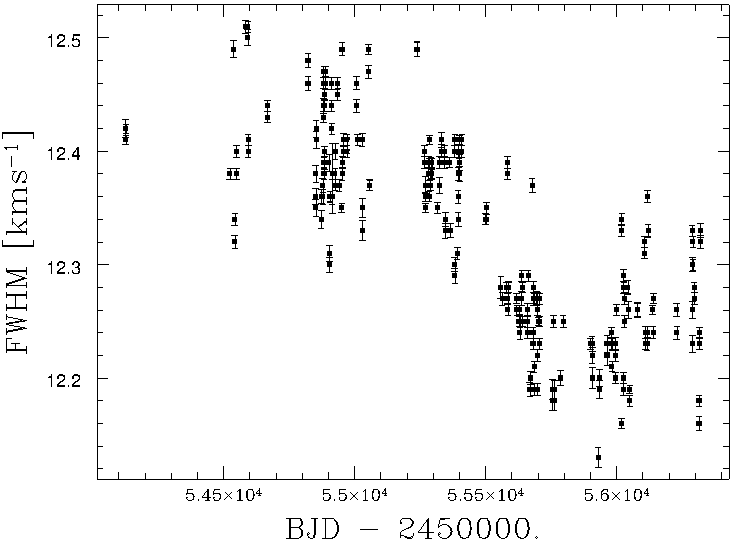}
     \includegraphics[width=0.8\hsize]{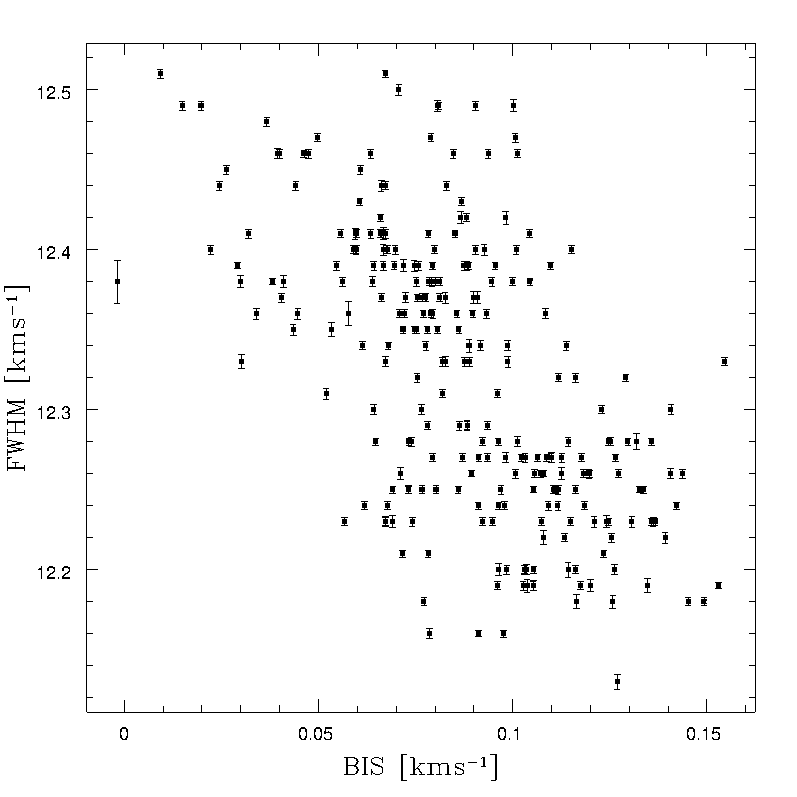}
\caption{{\it Top:} Temporal variations of the FWHM of the CCF. {\it Bottom:} FWHM versus BISs.}
    \label{FWHM}
  \end{figure}

% \subsection{CCF width }
The temporal variations of the FWHM of the CCF are reported in Figure~\ref{FWHM}. We do not see any variability with periods in the range 300 days. But we see a low amplitude long term variation. 
The FWHM values are well correlated with the BIS ones (see Fig.~\ref{FWHM}).
  
%==========================================================================
\section{Origins of the observed periodic RV variations}
%========================================================================
\label{keplerian_sol}

We observe both a 320-days periodic variability and a long term variability in the RV signal. 
We first note that given its very large projected separation, the M-type companions have no detectable impact on the spectrum and the RV variations of HD113337.
The weak long-term variations observed in the BIS and the FWHM of the CCFs are well correlated and present a similar long-term trend than the RV. This indicates that the long term signal is most probably not related to a gravitationally bound companion, but may be associated with long term stellar variability such as convection effects. In the following, we discuss the 320-days periodic RV variation.

\subsection{Stellar variability}

  Stellar pulsations are a very improbable origin as they would induce significant BIS variations \citep[see examples in][]{galland06} and because a $\geq$\,300 days periodicity is far larger than the ones of pulsations known for this type of main sequence stars.

Stellar spots can be excluded as they would lead to signals with periods of a few days or less. Assuming a radius of 1.5 \Rsun\,and a \vsini\,of 6.1 \kms, the rotation period would be less than 12 days. This is clearly not compatible with the observed $\simeq$320 days period. A mean log$R_{HK}'$ value of -4.8 is derived \citep{boisse10b}, which excludes a high level of activity and no variability is seen in the log$R_{HK}'$ data. Moreover, with this level of \vsini\ (\ie~$>$ 6 \kms), activity-induced RV variations would induce correlated BIS variations. 
The fact that no periodic signal is seen in the BIS at a period of 320 days allows us to dismiss stellar activity as an explanation to the 320 days variations \citep[see][]{desort07,boisse11}. We thus attribute them to the presence of a planetary companion. 

  %---------------------------------
  \subsection{A planet orbiting around HD113337}  
  \label{oneplanet}
  %---------------------------------
We fitted simultaneously the RV data of HD113337 with a Keplerian model and a quadratic law using a Levenberg-Marquardt algorithm, after selecting values with a genetic algorithm \citep{segransan11}. We treated the corrected \sophie\,and \sophiep\,data as independent samples. The best solution is an eccentric orbit ($e=0.46\pm0.04$) with a period of $P=324.0^{+1.7}_{-3.3}$ days and a semi-amplitude $K=75.6^{+3.7}_{-3.6}$ \ms. Taking into account the error bar on the stellar mass ($M_{\star}$=1.40  $\pm$0.14~M$_{\odot}$), the RV signature corresponds to a planet of minimum mass  $m_{P}\sin{i}$ = $2.83\pm 0.24$~\Mjup. The best-fit Keplerian model is plotted superimposed to the \sophie~and \sophiep~velocities in Fig~\ref{fit_2CC_all}. The final orbital elements are listed in Table~\ref{param_p}. They were computed using 5000 Monte Carlo simulations and the uncertainties in the final parameters correspond to their 1-sigma confidence intervals. The difference between the mean RV values from \sophie~and \sophiep~is consistent with observations from constant stars \citep{bouchy13}.  We checked that removing the first three data points does not change the parameters of the fit within 1-$\sigma$. 

The residuals are greater than the mean error bars, $\sim$25 and $\sim$19~\ms~for \sophie~and \sophiep, respectively. However, no periodic variation is detected. This high variability may originate both from an under-estimation of the instrument stability and from contaminant signal (\eg~moonlight).

 %--------------------------------------------------------------------------------
\renewcommand{\arraystretch}{1.25}% 1.25 -> 140% en plus
  \begin{table}[t!]
\caption{Best orbital solutions for HD113337 RVs.}
      \label{param_p}
     \begin{center}
        \begin{tabular}{l c }\\
  \hline
  \hline
          Parameter                                      & Values                          \\
  \hline
          $P$  [days]                                    & $324.0^{+1.7}_{-3.3} $             \\
          $T_0$   [JD$-$2400000]                         & $56074.5 \pm  2.3$              \\
          $e$                                            & $0.46 \pm  0.04$               \\
          $\omega$  [deg]                                & $-140.8 ^{+3.6}_{-3.7}$           \\
          $K$  [\ms]                                     & $75.6^{+3.7}_{-3.6}$              \\
  \hline
          linear    [\msy]                               & $-41.2 \pm 3.6$                \\
          quadratic  [\msysq]                            & $-15.9 \pm 0.6$                \\
  \hline
          $N_{\rm meas}$                                   & 266                             \\
          $\sigma_{O-C}$\textsc{sophie}  [\ms]            & 24.80 (60.45)~$^{\star}$          \\
          $\sigma_{O-C}$\textsc{sophie+}  [\ms]           & 18.82 (44.46)~$^{\star}$          \\
          $RV$$_{mean}$ \textsc{sophie}  [\,km\,s$^{-1}$]  &   $-0.039 \pm 0.005$            \\
          $RV$$_{mean}$ \textsc{sophie+}  [\,km\,s$^{-1}$] & $-0.021 \pm 0.006$              \\
          reduced $\chi^2$                               & $4.55$ ($11.60$)~$^{\star}$      \\
  \hline
          $m_{P}\sin{i}$ [\Mjup]                         & $2.83\pm 0.24$    ~$^{\dag}$     \\
          $a_{P}\sin{i}$  [AU]                           & $0.92 \pm 0.09$  ~$^{\dag}$      \\
  \hline

        \end{tabular}
      \end{center}
~$^{\star}$ The number in parenthesis refer to the model assuming a constant velocity.
$^{\dag}$ Assuming M$_{\star}$\,=\,1.40\,$\pm$\,0.14\,M$_{\odot}$ 
    \end{table}
    \renewcommand{\arraystretch}{1}% 1 -> 140% en plus
%-------------------------------------------------------------------------------  

\section{Discussion and concluding remarks}
Using our age estimation, HD113337 would be a particularly young planetary system detected by RV. Very few RV planets have yet been found around young stars. A giant planet of $6.1\pm 0.4$ \Mjup~was reported around the $\simeq$100 Myr-old G1V star HD70573 \citep{setiawan07}. More recently, \cite{vaneycken12} reported a possible close-in giant planet around a 7-10 Myr old T Tauri star, using both photometric and spectroscopic observations. Young stars being very active and rapidly rotating objects, one has to be particularly cautious when finding periodic RV variations which can be attributed to a planet. Indeed, stellar activity manifestations, such as cold spots, could mimic planet signatures, particularly with periodicity smaller or close to the stellar rotation period. There are some cases of RV signatures initially announced as a planet ones that later became controversial or were even rejected. An example is the signature of a $10 \pm 3$ \Mjup~planet reported by \cite{setiawan08} around TW Hya (8-10 Myr), which turned out to be the trace of a cold stellar spot, according to \cite{huelamo08} and to \cite{figueira10b}. Another case is the $\simeq 6.5\pm 0.5$ \Mjup~planet reported around the young (35-80 Myr) active K5V dwarf BD+201790 \citep{hernanobispo10}, which was also rejected later by \cite{figueira10a}. It is valuable to stress that in the case of HD113337, the reasons that lead to such false detections can be rejected: {\it i)} the 324-days period is far greater than the estimated stellar rotational period, {\it ii)} the level of stellar activity is much lower than for the mentioned cases, and {\it iii)} with a stellar \vsini~$>6$ \kms, RV variations induced by line profiles deformations would have been monitored by the BIS and FWHM parameters, which is not the case.

Interestingly, HD113337b properties are similar to those of the companion of 30 Ari B \citep{guenther09}, which has an orbital period and an excentricity very close to those of HD113337b, and a $\simeq 9.9$~\Mjup~minimum mass. 30 Ari B is a $\simeq 1.1$ \Msun~F4V star, with $T_{\rm eff} = 6462$ K and a higher \vsini~of $\simeq 38$~\kms. According to the age estimation made by these authors ($0.91 \pm 0.83$ Gyr), it is much likely older than HD113337. Whether the masses of 30 Ari Bb and HD113337b could be similar depends on the actual stars rotational velocity and on the inclination of the systems, both are unknown. We note that in the catalog of F-type dwarfs velocities of \cite{nordstroem97}, the mean \vsini~of 72 F4--F6 stars is about 49 \kms, while it is about 16 \kms~in our 49 F4--F6 stars sample. According to \cite{guenther09}, 30 Ari B rotational rate indicates that this star is probably seen almost equator-on, or with a small inclination. On the contrary, HD113337 \vsini~is significantly lower than the average values for the two mentionned samples. Combined with the fact that it is a young star, this could indicate a large inclination (and hence a low \vsini). Assuming an actual rotational velocity of 16 \kms~for HD113337 would lead to a $\simeq 20^{\circ}$\ inclination for our star. If the orbital and the stellar spin axes are aligned, 30 Ari Bb and HD113337b would then have almost the same mass. However, this is quite speculative at this stage.

  Young planetary systems are of peculiar interest as they can be targeted both by RV and by forthcoming deep imagers (such as SPHERE on the VLT). Detecting planets both in RV and imaging is a very important goal as it would give the opportunity to calibrate the brightness-mass relationships at young ages. These relationships are used to derive masses in imaging, and are still producing diverging results at young ages \citep[see \eg,][]{fortney08}. An example of the kind of constrains derived when combining RV and imaging on a yet more distant target can be found in \cite{lagrange12a}.

 Given its V-magnitude and declination, HD113337 is a good target for an interferometric instrument such as the VEGA spectrograph, operating at the Mount Wilson observatory. Using the largest baselines of the CHARA array, the star could be partially resolved, allowing one to determine accurately its angular diameter and to derive precise values of the stellar radius and mass \citep[the mass can be deduced thanks to the radius and surface gravity, see][]{ligi12}. Thus the Keplerian model of the planet could be better constrained. 
 An example of peculiar interest is the case of the F-type star, HD185395, for which a variability of the stellar angular diameter was detected \citep{ligi12}. This variability is still of unknown origin has the same periodicity than RV variations previously observed \citep{desort09}.

This system is also very interesting as it hosts at the same time a planet and a debris disk. From a statistical point of view, no correlation has been well defined between RV planets and IR excesses due to cold debris disks around solar-type stars. But it is not clear whether the absence of correlation can be due to evolution effects (the stars surveyed in RV are mature stars) or to other biases (the planets are at rather short separations). In a recent paper, \cite{wyatt12} suggested that systems composed of only small-mass planets ($m_{P}\sin{i}<$ Saturn mass) host also preferentially a debris disk. Although their result is subject to small number statistics, they proposed that this correlation could be a signature of  dynamically stable systems where planetesimals can remain unperturbed over Gyr timescale.
We also note that stars with imaged planets (at large separations) are surrounded by disks and are young. Many more systems like HD113337 are needed to further test the link between the debris disks and the presence of planets from an evolution point of view. Finally, high resolution imaging data of the inner part of the disk would be valuable to search for possible signs of disk-planet interactions. 

\begin{acknowledgements}
  We acknowledge support from the French CNRS and the support from the Agence Nationale de la Recherche (ANR grant XXX). We are grateful to the Observatoire de Haute-Provence (OHP) for their help during the observations, and to the Programme National de Plan\'etologie (PNP, INSU). These results have made use of the SIMBAD database, operated at CDS, Strasbourg, France. We also thank G\'erard Zins and Sylvain C\`etre for their help in implementing the SAFIR interface. NCS, IB and AS acknowledge the support by the European Research Council/European Community under the FP7 through Starting Grant agreement number 239953. NCS and IB also acknowledges the support from Funda\c{c}\~ao para a Ci\^encia e a Tecnologia (FCT) through program Ci\^encia\,2007 funded by FCT/MCTES (Portugal) and POPH/FSE (EC), and in the form of grants reference PTDC/CTE-AST/098528/2008 and PTDC/CTE-AST/098604/2008, and SFRH/BPD/81084/2011. 
AE is supported by a fellowship for advanced researchers from the Swiss National Science Foundation (grant PA00P2\_126150/1) and RFD is supported by CNES.
We gratefully acknowledge the comments and suggestions of the referee, A. Hatzes, that helped to improve the quality of the paper.
\end{acknowledgements}
\bibliographystyle{aa}
\bibliography{hd113337_biblio}
\end{document}